\documentclass[10pt, conference]{ieeeconf}
\IEEEoverridecommandlockouts

\usepackage[utf8]{inputenc}
\usepackage[T1]{fontenc}
\usepackage[english]{babel}
\usepackage[dvipsnames]{xcolor}
\usepackage{tabularx,hyperref,amsmath,amssymb,tikz,pgfplots,subcaption,circuitikz}

\let\labelindent\relax
\usepackage{amsthm}
\usepackage[inline]{enumitem}
\usetikzlibrary{calc,backgrounds,arrows.meta}
\pgfplotsset{compat=newest}
\pgfplotsset{every axis plot/.append style={line width=1pt,line join=round}}
\tikzset{every picture/.style={line width=1pt,>=latex}}
\tikzset{state/.style={outer sep=0pt,inner sep=0pt,draw,circle,minimum size=0.85cm}}

\belowdisplayskip \abovedisplayskip%
\DeclareMathOperator{\ac}{Ac}

\newtheorem{definition}{Definition}
\newtheorem{proposition}{Proposition}

\definecolor{mycolor1}{rgb}{0.00000,0.44700,0.74100}%

\newcommand{\compose}{\,\|\,}

\useshorthands{"}
\defineshorthand{"=}{\penalty1000\discretionary{-}{}{-}\penalty10000\hskip0pt}

\title{\LARGE \bf Discrete Event System Modeling of Neuromorphic Circuits}
\author{Koen~Scheres$^1$, Rodolphe~Sepulchre$^{1,2}$
\thanks{The research leading to these results has received funding from the European Research Council under the Advanced ERC Grant Agreement SpikyControl n.101054323. Email: {\itshape koen.scheres@kuleuven.be}, {\itshape rodolphe.sepulchre@kuleuven.be}.}%
\thanks{$^1$Department of Electrical Engineering (ESAT), KU Leuven, KasteelPark Arenberg 10, B-3001 Leuven, Belgium.}%
\thanks{$^2$Department of Engineering, University of Cambridge, Trumpington Street, Cambridge CB2 1PZ, United Kingdom.}%
}%

\begin{document}
\maketitle
\thispagestyle{plain}
\pagestyle{plain}

\begin{abstract}
Excitable neuromorphic circuits are physical models of event behaviors: their continuous-time trajectories consist of sequences of discrete events. This paper explores the possibility of extracting a discrete-event model out of the physical continuous-time model. We discuss the potential of this methodology for analysis and design of neuromorphic control systems.
\end{abstract}

\section{Introduction}
\begin{figure}[t!]
    \centering\small
    \begin{tikzpicture}
        \node[fill=CornflowerBlue!50, minimum width=.6\linewidth, minimum height=1cm, outer sep=0pt, text width=2cm, align=center] (A) at (0,0) {\textsc{Decision\\making}};
        \node[fill=Dandelion!50, minimum width=.6\linewidth, minimum height=1cm, outer sep=0pt, anchor=north, text width=2cm, align=center] (B) at (A.south) {\textsc{Trajectory\\planning}};
        \node[fill=LimeGreen!50, minimum width=.6\linewidth, minimum height=1cm, outer sep=0pt, anchor=north, text width=2cm, align=center] (C) at (B.south) {\textsc{Feedback\\control}};
        \filldraw[white, outer sep=0pt, inner sep=0pt, line join=round] (C.south west) -- (A.150) -- (A.north west) -- cycle;
        \filldraw[white, outer sep=0pt, inner sep=0pt, line join=round] (C.south east) -- (A.30) -- (A.north east) -- cycle;
        \draw[thick,line width=2pt,arrows={-Stealth[length=10pt]}] ($(C.south west)+(180:5mm)$) -- ($(A.north west)+(180:5mm)$) node[pos=0.5,above,rotate=90, text width=2.75cm, align=center, yshift=2mm] {Increased~flexibility\\and computation~time};
    \end{tikzpicture}
    \caption{By separating the control stack into a hierarchical structure, it is possible to do fast (real-time) tracking as well as slow decision-making. Adapted from \cite{Matni_Ames_Doyle_2024}.}
    \label{fig:hierarchy}
\end{figure}
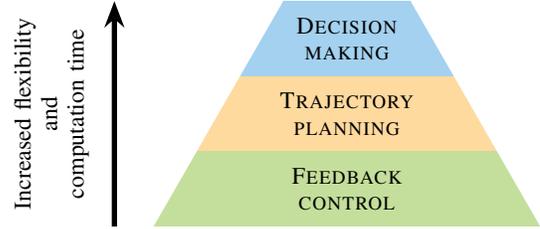

The majority of control and estimation problems can be separated in three categories: tracking, trajectory planning and decision-making. Although this separation is useful in many cases (e.g., to make analysis tractable), it necessitates a hierarchical control structure in the case where all are required. Indeed, in many high-tech (motion) control applications, first a low-level controller is designed that, e.g., tracks a given reference (and thus solves the tracking problem). Hereafter, a ``reference generator'' is designed which generates the reference for the low-level controller. Lastly, a ``supervisor'' is designed that typically handles the decision-making process (i.e., which reference to track) \cite{Matni_Ames_Doyle_2024}. This hierarchical structure is depicted in Fig.~\ref{fig:hierarchy}.
 
The different controllers are often designed (and potentially implemented) in different domains and/or devices. For example, the decision-making process is usually formulated in terms of automata, while the feedback control portion is typically tackled via differential equations. 
This division between the discrete and the continuous is not apparent in the animal world. Instead, central nervous systems (CNSs) regulate neuronal rhythms, made of events. At the cellular level, those events are made of ``spikes.'' Spikes are modeled by biophysical differential equations, yet they can be counted. Hence, they combine attributes of the continuous \emph{and} the discrete \cite{Sepulchre_2022}.

Many models of neurons are hybrid, but we are not aware of any methodological attempt to extract a discrete-event model from the biophysical model. However, leveraging the discrete nature allows us to endow neuromorphic circuits -- electronic circuits whose operating principles are inspired by neurons -- with discrete decision"=making capabilities. If we want to regard excitable systems as a unifying framework for regulation and automation, it seems crucial to combine continuous \emph{and} discrete descriptions of their behavior. The key advantage of such an approach is that the same architecture can be used throughout the hierarchical control stack. For example, for low-level control, we can exploit \emph{central pattern generators}, which are essential building blocks in the CNS for locomotion \cite{Marder_Bucher_2001}. On the other hand, for high-level control we can use the \emph{winner-take-all} principle, which was pioneered by Hopfield \cite{Hopfield_1984} in the classical Hopfield networks, and later shown to enable interesting computation capabilities and even universal approximation properties in some cases \cite{Maass_1999}.

This background motivates the question of the present paper: what is the discrete-event model of a continuous-time biophysical model of excitable neurons? The answer to this question allows us to systematically solve verification problems (which may be phrased, e.g., as temporal logic) or exploit other tools available in the discrete event system literature, but instead applied to neuromorphic circuits.

The rest of this paper is structured as follows: In Section~\ref{sec:des}, we will recall some basic concepts from discrete event systems. Next, we will explain the basic properties and behaviors of neurons and neuromorphic circuits in Section~\ref{sec:neuro}. In Section~\ref{sec:des_neuro}, we will give several models in terms of automata for different behaviors potentially present in neurons. Next, we will delineate how to construct decision-making circuits and how this corresponds ``one-to-one'' with automata descriptions. We illustrate the latter by shortly discussing a realization theory for automata as neuromorphic circuits. Lastly, we will summarize and discuss our findings in Section~\ref{sec:conclusions}.

\section{Discrete Event Systems}\label{sec:des}
This section provides a short recap on Discrete Event Systems (DESs) and different types of modeling classes that are of interest in modeling and simulating neuromorphic circuits. 
\begin{definition}[{\cite[p. 30]{Cassandras_Lafortune_2021}}]\label{def:des}
    A \emph{Discrete Event System} is a discrete-state, event-driven system, that is, its (state) evolution depends entirely on the occurrence of asynchronous discrete events over time.
\end{definition}

Many model classes exist within the DES literature. In the context of the present paper, we only consider two (sub)classes, namely, \emph{timed} and \emph{untimed} automata. In case of the untimed automaton, the interest lies in the \emph{ordering} of the events. Thus, untimed automata can be fully characterized in terms of their (marked) language, with the following formal definition.
\begin{definition}[{\cite[p. 55]{Cassandras_Lafortune_2021}}]\label{def:language}
    A \emph{language} defined over an event set $E$ is a set of finite-length strings formed from events in~$E$.
\end{definition}
The language of an automaton thus restricts the possible sequences of events that can occur. In this paper, we only consider regular languages, that is, languages that can be generated by a finite-state automaton.

We refer to \cite{Cassandras_Lafortune_2021} for classical notions such as the accessible part $\ac(A)$ of automaton $A$ and the composition operator $A\compose B$ that ``joins'' the automata $A$ and~$B$.

When timing information is explicitly considered, such as in timed automata, the sample paths of the system are determined by both the event and its timing, i.e., by an ordered sequence $\{(e_1,t_1),(e_2,t_2),\ldots\}$. Although it may be relevant to analyze event timings in neuromorphic circuits, as a first step we are restricting ourselves to \emph{untimed} automata in this paper.

\section{Neuromorphic Circuits}\label{sec:neuro}
This section summarizes key behavioral properties of neurons, synapses, and neuromorphic circuits.
\subsection{The event behavior of an excitable neuron}
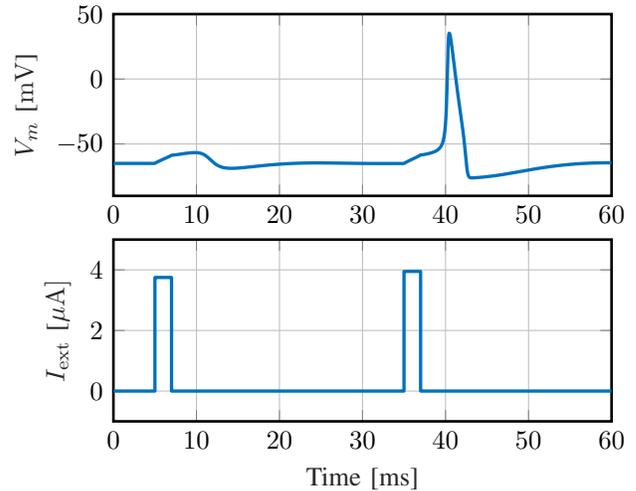
\begin{figure}[t!]
    \centering
    \begin{tikzpicture}

\begin{axis}[%
width=0.95\linewidth,
height=4cm,
at={(0,0)},
% scale only axis,
xmin=0,
xmax=60,
ymin=-90,
ymax=50,
ylabel style={font=\color{white!15!black}},
ylabel={$V_m$ [$\mathrm{mV}$]},
axis background/.style={fill=white},
xmajorgrids,
ymajorgrids
]
\addplot [color=mycolor1, line width=1.25pt, forget plot]
  table[row sep=crcr]{%
0	-65\\
4.89275998486926	-65.0001318527616\\
4.92760915437717	-65.0154650314394\\
4.96245832388507	-64.9958922822339\\
4.99730749339298	-64.9002399762916\\
5.13670417142461	-64.3978653417861\\
5.39602887522511	-63.5578900402474\\
5.71967790763203	-62.5929595757266\\
6.7316438975471	-59.6632043179476\\
6.87234615128504	-59.1417871969737\\
7.01304840502297	-58.7277457948516\\
7.1537506587609	-58.5812293951523\\
7.29445291249883	-58.5481270489933\\
7.46706261653755	-58.4055263921259\\
8.6841888780578	-57.2854813019817\\
9.02524027882083	-57.0227978633033\\
9.36629167958387	-56.8035237065316\\
9.53681737996538	-56.7252824273469\\
9.7073430803469	-56.6905349906784\\
9.87786878072842	-56.6884892273796\\
10.0175348686505	-56.7055679786402\\
10.1572009565726	-56.7697507675452\\
10.2968670444947	-56.8954306087713\\
10.4365331324168	-57.0765242684319\\
10.553133101718	-57.2697497155254\\
10.6697330710192	-57.5193679830042\\
10.7863330403204	-57.8312941565258\\
10.9029330096216	-58.2060452028858\\
11.0300550330764	-58.6867137933327\\
11.1571770565312	-59.2446676519583\\
11.4114211034407	-60.5632353674078\\
12.1497981319154	-64.7987398750339\\
12.3069543827133	-65.5479714185552\\
12.4641106335112	-66.2038980024801\\
12.6212668843091	-66.7652388067599\\
12.778423135107	-67.2370010383616\\
12.9668089736274	-67.694660134318\\
13.1551948121478	-68.049903573445\\
13.3401912002147	-68.3132220111432\\
13.5251875882817	-68.5057571649001\\
13.7269208267612	-68.6484121256811\\
13.9286540652407	-68.7336822558067\\
14.0433639106609	-68.7606381578987\\
14.2727836015014	-68.7759440169883\\
14.52279390948	-68.7447494649546\\
14.7933948345965	-68.6678067799385\\
15.0914931486593	-68.5442665275619\\
15.4170888516686	-68.376150411653\\
15.7603439848157	-68.173204550112\\
16.3017158297435	-67.8226379692171\\
17.9848289540836	-66.699481918878\\
18.6198866031974	-66.3122907725049\\
19.1308168076126	-66.025067185172\\
19.650435251672	-65.7585680222545\\
20.165484302965	-65.5216166641799\\
20.6637809023217	-65.3182943221735\\
21.1643619803027	-65.1409774643007\\
21.6740421409653	-64.988865333253\\
22.199635988367	-64.8618112698353\\
22.7459453933592	-64.7602453930973\\
23.2909779279385	-64.6888220498735\\
23.8294046951801	-64.6465559640215\\
24.3558967981592	-64.6294183178502\\
24.8719570833517	-64.6317356274975\\
25.5698746735473	-64.6606970267992\\
26.47219424971	-64.7283247569291\\
29.9951964670321	-65.0178747829879\\
31.0907063151688	-65.0595722973612\\
32.1439828284017	-65.0765124661154\\
33.702271379347	-65.0687859063928\\
34.9869106879002	-65.0487099395197\\
35.025868112533	-64.9660080697768\\
35.1816978110644	-64.3687935503674\\
35.457853848451	-63.4280043178879\\
35.7952852946107	-62.3698572374339\\
36.5542112292578	-60.0519222074965\\
36.8159734324018	-59.1741269725625\\
37.0069748211203	-58.4671022757367\\
37.0370349646935	-58.4342003481944\\
37.0971552518399	-58.4250965789959\\
37.1873356825596	-58.372115081376\\
37.3440513745772	-58.2322235760104\\
37.5825060151877	-57.9487366520547\\
37.8277228609665	-57.6108014162536\\
38.0512230264268	-57.265699041038\\
38.274723191887	-56.8715449998604\\
38.4334583725395	-56.5499330759248\\
38.5921935531919	-56.1826208746187\\
38.7509287338444	-55.7529532100438\\
38.9096639144968	-55.2283089834524\\
39.0535038701935	-54.6250620151187\\
39.1973438258901	-53.8702761217928\\
39.3411837815868	-52.8836882561932\\
39.4850237372834	-51.4957704068347\\
39.5678520055867	-50.4140883304926\\
39.65068027389	-49.0379462482518\\
39.7335085421933	-47.2224514244695\\
39.8163368104966	-44.732329714726\\
39.9159304793411	-40.2913952300189\\
39.9657273137634	-37.1133834307384\\
40.0155241481856	-32.958077731248\\
40.0653209826079	-27.4965882912017\\
40.1151178170301	-20.0757087209945\\
40.2147114858746	1.61854481511712\\
40.3181389757135	25.9629240503153\\
40.3675066760768	32.3048267300795\\
40.4045324513492	34.5452437582922\\
40.4303251122771	35.2061754104965\\
40.4572265839511	35.3605993099327\\
40.4848548122645	35.1366171429742\\
40.5273872896937	34.296665502725\\
40.5932273902255	32.2954011258498\\
40.6868627897015	28.5959269034071\\
40.8352557686062	21.5589821341211\\
41.1772094117322	3.69980602348869\\
41.4932825256368	-12.0533025089954\\
41.8176543363438	-26.9061516612488\\
42.1097743465728	-40.4147116093327\\
42.2379923144215	-47.2926543972015\\
42.3669735876394	-55.329178672367\\
42.5665278264638	-68.1470281060457\\
42.6347280031986	-71.2829179770442\\
42.7011870012953	-73.3650260077073\\
42.7659048207538	-74.6206028294692\\
42.8363867916362	-75.3951588967399\\
42.9126329139425	-75.8211569107508\\
42.9817369530845	-76.0068917370266\\
43.0436989090624	-76.0914438970945\\
43.1045199448488	-76.1301092702167\\
43.1940401182415	-76.1441357972232\\
43.3205780621369	-76.1226135595573\\
43.5019800670469	-76.0593451766975\\
43.8207204267291	-75.9168482902231\\
44.2419847538082	-75.6968342108975\\
44.6439593312515	-75.4566671558684\\
45.0637792775925	-75.1729989398602\\
45.5217914327285	-74.8266351868164\\
45.9135544935399	-74.5001734128833\\
46.3523049932796	-74.1033358039517\\
46.8203354897971	-73.6470724820044\\
47.4332913864317	-73.0063388846545\\
48.1362120125232	-72.2276591129443\\
50.6923770994451	-69.339637327896\\
51.371488416468	-68.6339194975019\\
51.9300346547288	-68.0885998616159\\
52.5088089201138	-67.5607059860623\\
53.1059541924611	-67.0581945999929\\
53.5630027854495	-66.7037074421758\\
54.0324621697876	-66.3668487550591\\
54.5107562909357	-66.0525967469453\\
54.9977673607655	-65.7632928530845\\
55.4933775911482	-65.5007847478942\\
56.000144686215	-65.2650318076245\\
56.5128477542659	-65.060521923568\\
57.0308202183973	-64.8891390331193\\
57.5533955017056	-64.751512110915\\
58.0812623718479	-64.6465297776433\\
58.6116343351968	-64.5745132276269\\
59.1439375909049	-64.5344394879735\\
59.6775983381245	-64.5232703026954\\
60	-64.5282074217069\\
};
\end{axis}

\begin{axis}[%
width=0.95\linewidth,
height=4cm,
at={(0cm,-3cm)},
% scale only axis,
xmin=0,
xmax=60,
xlabel style={font=\color{white!15!black}},
xlabel={Time [ms]},
ymin=-1,
ymax=5,
ylabel style={font=\color{white!15!black}},
ylabel={$I_{\mathrm{ext}}$ [$\mathrm{\mu A}]$},
axis background/.style={fill=white},
xmajorgrids,
ymajorgrids
]
\addplot [color=mycolor1, line width=1.25pt, forget plot]
  table[row sep=crcr]{%
0	0\\
5	0\\
5	3.75\\
7	3.75\\
7	0\\
35	0\\
35	3.95\\
37	3.95\\
37	0\\
60	0\\
};
\end{axis}
\end{tikzpicture}%
    \caption{A neuron is an \emph{excitable} system: a minor variation in the input can cause a major variation in the output due to the presence of a bifurcation.}\label{fig:excitability}
\end{figure}

The ``discreteness'' of such excitable behaviors is illustrated in Fig. \ref{fig:excitability}: it stems from the existence of a threshold, which itself stems from a mixed pair of active currents. Each threshold accounts for a novel discrete state. A neuron with two thresholds, one for spiking and one for bursting, will have two discrete states that can be distinguished from the resting state.

\begin{figure}[t!]
    \centering
    \begin{circuitikz}[american voltages]
        \node[ocirc] (posTerm) at (0,0) {};
        \node[ocirc] (negTerm) at (0,3cm) {};
        \path[draw] (posTerm) to ($(posTerm)+(0:1cm)$) to[capacitor, l={$C$}] ++(90:3cm) -| (negTerm.east);

        \path[draw] (posTerm) to ++(0:2cm) to[battery1,invert] ++(90:0.75cm) to[american resistor] ++(90:1.5cm) to ++(90:0.75cm) -| (negTerm.east);
        \path[draw] (posTerm) to ++(0:3.25cm) to[battery1,invert] ++(90:0.75cm) to[variable american resistor,mirror] ++(90:1.5cm) to ++(90:0.75cm) edge[->] node[midway, left] {$I_{\mathrm{Na}}$} ++(-90:0.75cm) -| (negTerm.east);
        \path[draw] (posTerm) to ++(0:4.5cm) to[battery1,invert] ++(90:0.75cm) to[variable american resistor,mirror] ++(90:1.5cm) to ++(90:0.75cm) edge[->] node[midway, left] {$I_{\mathrm{K}}$} ++(-90:0.75cm) -| (negTerm.east);
        \draw[<->] (-0.75,0) -- (-0.75, 3) node[midway, left] {$V$} node[pos=0, right] {$-$} node[pos=1, right] {$+$};
        \draw[->] (negTerm) -- ++ (0:1cm) node[midway, above] {$I$};
    \end{circuitikz}
    \caption{Circuit representation of Hodgkin-Huxley model.}\label{fig:circuit}
\end{figure}
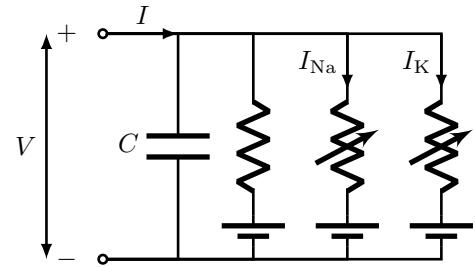

The biophysical modeling principles of neuronal modeling go back to the seminal work of Hodgkin and Huxley \cite{Hodgkin_Huxley_1952}. The neuron is modeled as an electrical circuit with the specific architecture shown in Fig. \ref{fig:circuit}: an RC circuit (modeling the passive membrane) connected to a parallel bank of current sources, each made of a voltage-dependent conductor in series with a constant battery.

\subsection{Excitatory and inhibitory synaptic currents}\label{subsec:synapses}
The inputs to neurons are synaptic currents. In line with the previous section, we can categorically separate these current in two types: \emph{inhibitory} currents and \emph{excitatory} currents. An inhibitory current hyperpolarizes the membrane, which ``raises'' distance to the threshold; this leads to the neuron being less sensitive to inputs. An excitatory current, however, depolarizes the membrane, which ``lowers'' the distance to the threshold, bringing the membrane potential closer to (or over) the threshold. 

\begin{figure}[t!]
    \centering
    \input{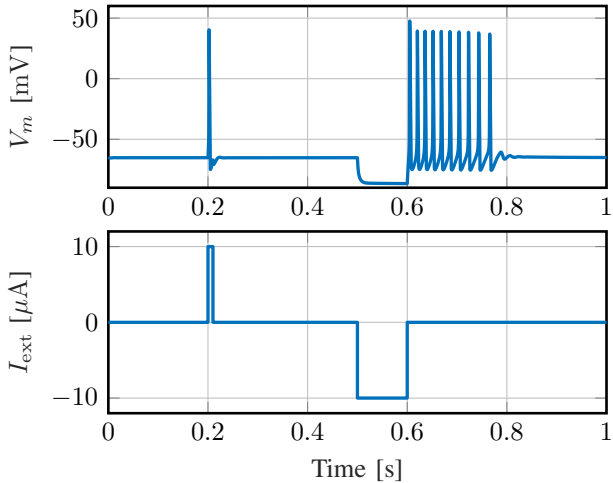}
    \caption{Different responses to stimuli: excitatory spiking and post-inhibitory bursting of the Hodgkin-Huxley model with additional $T$-type $\mathrm{{Ca}^{2+}}$ and slow $\mathrm{K}^+$ ion channels.}\label{fig:spiking_and_bursting}
\end{figure}

A neuron whose membrane is inhibited (i.e., hyperpolarized) for a sufficiently long time may, depending on the included ion channels and parameters, display the so-called \emph{post-inhibitory rebound spiking or bursting} phenomena. Whether a neuron is post-inhibitory rebound spiking or bursting depends on which ion channels are included in the model. This phenomenon is illustrated in Fig. \ref{fig:spiking_and_bursting}.
It is well-known that both inhibition and excitation are required for decision-making processes \cite{Hahnloser_Sarpeshkar_Mahowald_Douglas_Seung_2000}, hence we deem this distinction significant. We will illustrate this shortly in Section~\ref{sec:wta}.

\subsection{Event triggering by external inputs}\label{subsec:inputs}
To create meaningful DES models, we have to include a \emph{transition function}, i.e., a partial function that maps states and events to states. If no external current is supplied to a neuron in its resting state, it will never pass the threshold, and thus remain in its resting state. However, when an external current is supplied, a neuron will \emph{always} return to its resting state endogenously, i.e., \emph{regardless of} external inputs. This important distinction is not present in classical DES literature. In a DES, events are typically separated into controllable and uncontrollable events, wherein a ``supervisor'' can disable specific controllable events. 
Instead, we need a distinction between external and internal transitions. To that end, we borrow the concept of event forcing in \cite{Reniers_Cai_2025}.

Such a ``hierarchy'' in the ordering of events is natural in neuromorphic circuits: we can separate the ``endogenous'' or \emph{internal} behavior from the ``exogenous'' or \emph{external} behavior. The concept of \emph{event forcing} can thus be used to enrich the solution space of a DES; essentially, we enforce that, in the presence of an input, all external events preempt internal ones. Lastly, as both excitatory and inhibitory currents may enable distinct transitions (such as in Fig. \ref{fig:spiking_and_bursting}), it is of interest to label a transition as resulting from either an excitatory or an inhibitory input.

\subsection{Separating the discrete from the continuous}
We are interested in DES abstractions of the continuous-time differential equation of the conductance-based models, that only model the discrete, that is: \begin{enumerate*}[label=(\roman*)]\item the discrete states, which are in one to one correspondence with the \emph{pairs of mixed conductances}; \item the transitions between the discrete states, which are either internal or external; and, \item the discrete classification between excitation and inhibition.\end{enumerate*}

The remaining properties of the conductance-based model are regarded as ``continuous''. The role of the discrete-event model is to extract from the continuous-time differential equation what is discrete and \emph{only} what is discrete. All other properties belong to the continuous-time model. For example, the timing of events is regarded as a continuous property, as it will be continuously regulated by neuromodulators, which can be modeled as continuous parameter variation in the differential equation. This also clearly illustrates why modeling a neuromorphic circuit as a DES is a complement rather than a replacement of the physical model: we wish to use the DES to investigate the temporal ordering of events (e.g., the decision-making), while we want to resort to the conductance-based differential equation to study the continuous-time behavior (i.e., the regulation problem or feedback control).

In line with the above, the modeling abstractions we introduce below hold for neuronal models that exhibit certain \emph{behaviors}. We want a systematic methodology to associate a specific DES model to a specific biophysical circuit model. We will therefore insist on the following conventions in the construction of DES models.
\begin{definition}\label{ass:des-rules}
The DES model of a neuromorphic circuit will adhere to the following rules:
    \begin{enumerate}
        \item Each neuron has a ``resting'' or ``idle'' state;
        \item Each threshold requires a mixed pair of conductances, and, consequently, each threshold corresponds to a distinct state;
        \item Events are asynchronous, i.e., events that always occur simultaneously will be considered as a singular event;
        \item Transitions are separated in \emph{internal} and \emph{external};
        \item Internal transitions are denoted by {\protect\tikz[baseline=-0.5ex]{\protect\draw[->,anchor=base](0,0)--(0:8mm);}}. External transitions are separated in \emph{excitatory}, i.e., depolarizing, and \emph{inhibitory}, i.e., hyperpolarizing. They are denoted by {\protect\tikz[baseline=-0.5ex]{\protect\draw[{Turned Square}->,densely dashed,anchor=base](0,0)--(0:8mm);}} and {\protect\tikz[baseline=-0.5ex]{\protect\draw[{Circle[open]}->,densely dashed,anchor=base](0,0)--(0:8mm);}}, respectively;
        \item External transitions \emph{preempt} internal transitions.
    \end{enumerate}
\end{definition}

\section{Discrete Event System models of Neuromorphic Circuits}\label{sec:des_neuro}
In this section, we apply the above principles to a number of well-known neuromorphic ``motifs.''
\subsection{DES models of single neurons}
\begin{figure}[t!]
    \centering
    \begin{subfigure}[b]{0.26\linewidth}
        \centering
        \begin{tikzpicture}
            \node[state] (idle) at (0,0) {$i$};
            \draw[<-] (idle) -- ++(90:1cm);
            \path[{Turned Square}->,densely dashed] (idle) edge[out=20, in=-20, loop] node[midway, right] {$\sigma$} (idle);
        \end{tikzpicture}
        \caption{LIF neuron}\label{fig:lif}
    \end{subfigure}~~
    \begin{subfigure}[b]{0.33\linewidth}
        \centering
        \begin{tikzpicture}
            \node[state] (idle) at (0,0) {$i$};
            \node[state] (spiking) at ($(idle)+(0:1.75cm)$) {$s$};
            \draw[<-] (idle) -- ++(90:1cm); 
            \path[{Turned Square}->,densely dashed] (idle) edge[bend right=5mm] node[midway, below] {$\sigma$} (spiking);
            \path[->] (spiking) edge[bend right=5mm] node[midway, above] {$\eta$} (idle);
        \end{tikzpicture}
        \caption{Hodgkin-Huxley}\label{fig:excitable-neuron}
    \end{subfigure}~~
    \begin{subfigure}[b]{0.33\linewidth}
        \centering
        \begin{tikzpicture}
            \node[state] (idle) at (0,0) {$i$};
            \node[state] (spiking) at ($(idle)+(0:1.75cm)$) {$s$};
            \draw[<-] (idle) -- ++(90:1cm); 
            \path[{Turned Square}->,densely dashed] (idle) edge[] node[midway, above] {$\sigma$} (spiking);
            \path[{Circle[open]}->,densely dashed] (idle.-60) edge[bend right=5mm] node[midway, above] {$\varrho$} (spiking.-120);
            \path[->] (spiking.120) edge[bend right=6mm] node[midway, above] {$\eta$} (idle.60);
        \end{tikzpicture}
        \caption{Rebound spiking}\label{fig:rebound-spiking-neuron}
    \end{subfigure}
    \caption{Automaton representations of single spiking neurons.}
\end{figure}
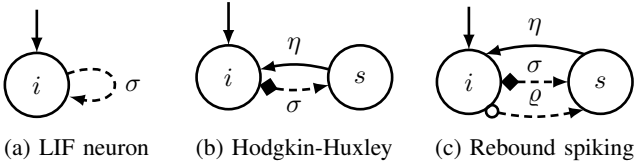

What is the DES of a single neuron? The simplest neuronal model is the leaky integrate-and-fire neuronal model, modeled as a leaking integrator with reset:
\begin{equation}
    \begin{aligned}
        \dot{x}(t)&=-\alpha x+u, & x\leqslant\theta,\\
        x(t^+)&=0, & x>\theta,
    \end{aligned}
\end{equation}
where $\alpha>0$ denotes the membrane leakage and $\theta>0$ the threshold. 
This model lacks the mixed pair of conductances of the biophysical model. The excitability threshold is only modeled as a reset phenomenon. Thus, in this simple model, the reset is identified with the event. As the model has no threshold resulting from a balance between mixed conductances, the associated discrete-event model only has one state, namely the idle state, and a single transition, which is a self-loop, see Fig.~\ref{fig:lif}.

In contrast, the DES model extracted from the Hogkin-Huxley circuit has a discrete state distinct from the resting state. The transition from rest to spike is external, because it requires an external input. In contrast, the transition from spike to rest is internal. This results in the automaton depicted in Fig.~\ref{fig:excitable-neuron}.

The model in Fig.~\ref{fig:rebound-spiking-neuron} is a further refinement of the model in Fig. \ref{fig:excitable-neuron}: it captures the additional rebound property of Hodgkin-Huxley model. As both excitatory and inhibitory inputs can trigger a spike, the full model requires an extra transition from the \emph{idle} state to the \emph{spiking} state. Both of these transitions require external inputs in the single neuron case, however, one is induced by excitation and one by inhibition. The complete model does not require an additional state, however, as only one pair of mixed conductances is present. 

The spiking neuronal model can be generalized to a bursting neuronal model. In a bursting model, the circuit includes an additional pair of mixed conductances, that creates a ``bursting'' threshold in addition to the ``spiking'' threshold. Consequently, a bursting neuron has three rather than two states: idle, spiking, and bursting, or $i$, $s$, and $b$, respectively. Depending on the neuron parameters, a burst event can result from excitation (burst excitability), from inhibition (rebound bursting), or both. Depending on the input trigger, the neuron can either spike or burst. All these models share the same set of states but different transitions from the idle state to the spiking or bursting states. Note that it may be impossible to reach the spiking state via external currents in some models.

\begin{figure}[t!]
    \centering
    \begin{tikzpicture}
        \node[state] (idle) at (0,0) {$i$};
        \node[state] (spiking) at ($(idle)+(0:1.75cm)$) {$s$};
        \node[state] (bursting) at ($(idle)-(0:1.75cm)$) {$b$};
        \draw[<-] (idle) -- ++(90:1cm); 
        \path[{Turned Square}->,densely dashed] (idle) edge[bend right=5mm] node[midway, below] {$\sigma$} (spiking);
        \path[{Circle[open]}->,densely dashed] (idle) edge[bend right=5mm] node[midway, above] {$\beta$} (bursting);
        \path[->] (spiking) edge[bend right=5mm] node[midway, above] {$\eta$} (idle);
        \path[->] (bursting) edge[bend right=5mm] node[midway, below] {$\rho$} (idle);
    \end{tikzpicture}
    \caption{Automaton representation of a spiking and rebound bursting neuron.}\label{fig:rebound-bursting-neuron}
\end{figure}
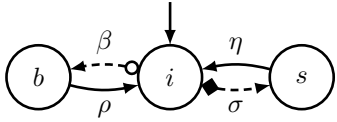

A common and important phenomenon in biological neurons is a neuron that spikes under depolarization (i.e., when an excitatory current is applied), but rebound bursts under hyperpolarization (i.e., when a sufficiently long inhibitory current is applied). This is the behavior modeled in Fig.~\ref{fig:spiking_and_bursting}. As we will refer to this neuron later, we explicitly provide the DES for this particular model, which is depicted in Fig.~\ref{fig:rebound-bursting-neuron}.

\subsection{DES models of synaptic interconnections}
We now add synapses to interconnect several neurons. To differentiate between the different states and events of the different neurons, we index the states by the number of the neuron, e.g., $s_i$ and $\sigma_i$ belong to the $i$-th neuron. As explained in Section~\ref{subsec:synapses}, synapses come in two types, which we treat separately in the following.

In general, any synaptic connection between two neurons will turn an \emph{external} transition into an \emph{internal} transition. Depending on the synapse type and the neuron it is connected to, it will add a new \emph{synchronized} internal transition on top of either the preexisting excitatory external transition or the preexisting inhibitory external transition. The key here is that these new transitions are \emph{synchronized to events/transitions in other neurons}. As a consequence, when there is a synaptic connection from neuron $\mathcal{N}_1$ to neuron $\mathcal{N}_2$, the accessible state-space in the composed automaton $\mathcal{N}_1\compose\mathcal{N}_2$ is \emph{reduced}. In Willem's behavioral terminology, a synapse \emph{restricts} the behavior by variable sharing, that is, transition synchrony.

In all the examples given below, we use rebound spiking neurons (see Fig. \ref{fig:rebound-spiking-neuron}) and assume that we only interact with neuron $\mathcal{N}_1$, i.e., we prune the external transitions from neuron~$\mathcal{N}_2$.

\subsubsection{Excitatory synapse}
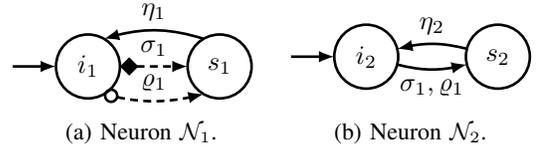
\begin{figure}[t!]
    \centering
    \begin{subfigure}{0.4\linewidth}
        \begin{tikzpicture}
            \node[state] (idle) at (0,0) {$i_1$};
            \node[state] (spiking) at ($(idle)+(0:1.75cm)$) {$s_1$};
            \draw[<-] (idle) -- ++(180:1cm); 
            \path[{Turned Square}->,densely dashed] (idle) edge[] node[midway, above] {$\sigma_1$} (spiking);
            \path[{Circle[open]}->,densely dashed] (idle.-60) edge[bend right=5mm] node[midway, above] {$\varrho_1$} (spiking.-120);
            \path[->] (spiking.120) edge[bend right=6mm] node[midway, above] {$\eta_1$} (idle.60);
        \end{tikzpicture}
        \caption{Neuron $\mathcal{N}_1$.}
    \end{subfigure}
    \begin{subfigure}{0.4\linewidth}
        \centering
        \begin{tikzpicture}
            \node[state] (idle2) at (0,0) {$i_2$};
            \node[state] (spiking2) at ($(idle2)+(0:1.75cm)$) {$s_2$};
            \draw[<-] (idle2) -- ++(180:1cm);
            \path[->] (idle2) edge[bend right=5mm] node[midway, below] {$\sigma_1,\varrho_1$} (spiking2);
            \path[->] (spiking2) edge[bend right=5mm] node[midway, above] {$\eta_2$} (idle2);
        \end{tikzpicture}
        \caption{Neuron $\mathcal{N}_2$.}
    \end{subfigure}
    \\[\belowdisplayshortskip]
    \begin{subfigure}[b]{0.55\linewidth}
        \centering
        \begin{tikzpicture}
            \node[state] (idle) at (0,0) {$i_1i_2$};
            \node[state] (spiking) at ($(idle)+(0:1.75cm)$) {$s_1s_2$};
            \node[state] (idle_spiking) at ($(spiking)+(-30:1.5)$) {$i_1s_2$};
            \node[state] (spiking_idle) at ($(spiking)+(30:1.5)$) {$s_1i_2$};
            \draw[<-] (idle) -- ++(90:1cm);
            \path[{Turned Square}->,densely dashed] (idle) edge[bend right=5mm] node[midway, below] {$\sigma_1$} (spiking);
            \path[{Circle[open]}->,densely dashed] (idle) edge[bend left=5mm] node[midway, above] {$\varrho_1$} (spiking);
            \path[->] (spiking) edge node[pos=0.4, below] {$\eta_1$} (idle_spiking);
            \path[->] (spiking) edge node[pos=0.4, above] {$\eta_2$} (spiking_idle);
            \path[->] (spiking_idle.150) edge[bend right=6mm] node[pos=0.4,above] {$\eta_1$} (idle.60);
            \path[->] (idle_spiking.-150) edge[bend left=6mm] node[pos=0.4,below] {$\eta_2$} (idle.-60);
        \end{tikzpicture}
        \caption{$\mathcal{N}_1\compose\mathcal{N}_2$ with $\mathcal{N}_1\stackrel{e}{\mapsto}\mathcal{N}_2$.}\label{fig:composed-excitatory}
    \end{subfigure}
    \caption{Automata with excitatory synapse: the external event $\sigma_2$ is replaced by the internal events $\sigma_1$ and $\varrho_1$.}\label{fig:automaton-excitatory-synapse}
\end{figure}

If there is an excitatory synaptic connection from neuron $\mathcal{N}_1$ to neuron $\mathcal{N}_2$, compactly denoted as $\mathcal{N}_1\stackrel{e}{\mapsto}\mathcal{N}_2$, then $\sigma_2$ becomes an internal event synchronized with $\sigma_1$ \emph{and} $\varrho_1$. The resulting automata are given in Fig. \ref{fig:automaton-excitatory-synapse}.

\subsubsection{Inhibitory synapse}
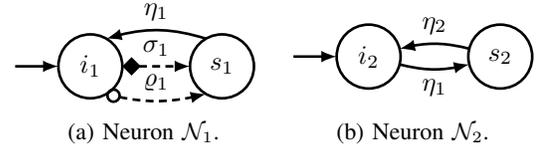
\begin{figure}[t!]
    \centering
    \begin{subfigure}{0.4\linewidth}
        \begin{tikzpicture}
            \node[state] (idle) at (0,0) {$i_1$};
            \node[state] (spiking) at ($(idle)+(0:1.75cm)$) {$s_1$};
            \draw[<-] (idle) -- ++(180:1cm); 
            \path[{Turned Square}->,densely dashed] (idle) edge[] node[midway, above] {$\sigma_1$} (spiking);
            \path[{Circle[open]}->,densely dashed] (idle.-60) edge[bend right=5mm] node[midway, above] {$\varrho_1$} (spiking.-120);
            \path[->] (spiking.120) edge[bend right=6mm] node[midway, above] {$\eta_1$} (idle.60);
        \end{tikzpicture}
        \caption{Neuron $\mathcal{N}_1$.}
    \end{subfigure}
    % \\[\belowdisplayshortskip]
    \begin{subfigure}{0.4\linewidth}
        \centering
        \begin{tikzpicture}
            \node[state] (idle2) at (0,0) {$i_2$};
            \node[state] (spiking2) at ($(idle2)+(0:1.75cm)$) {$s_2$};
            \draw[<-] (idle2) -- ++(180:1cm);
            \path[->] (idle2) edge[bend right=5mm] node[midway, below] {$\eta_1$} (spiking2);
            \path[->] (spiking2) edge[bend right=5mm] node[midway, above] {$\eta_2$} (idle2);
        \end{tikzpicture}
        \caption{Neuron $\mathcal{N}_2$.}
    \end{subfigure}
    ~
    \\[\belowdisplayshortskip]
    \begin{subfigure}[b]{0.45\linewidth}
        \centering
        \begin{tikzpicture}
            \node[state] (idle) at (0,0) {$i_1i_2$};
            \node[state] (spiking_idle) at ($(idle)+(30:1.75cm)$) {$s_1i_2$};
            \node[state] (idle_spiking) at ($(idle)+(-30:1.75)$) {$i_1s_2$};
            \draw[<-] (idle) -- ++(180:1cm); 
            \path[{Turned Square}->,densely dashed] (idle) edge[bend right=5mm] node[pos=0.4,below,anchor=north west] {$\sigma_1$} (spiking_idle);
            \path[{Circle[open]}->,densely dashed] (idle) edge[bend left=5mm] node[pos=0.6, above,anchor=south east] {$\varrho_1$} (spiking_idle);
            \path[->] (spiking_idle) edge node[midway, right] {$\eta_1$} (idle_spiking);
            \path[->] (idle_spiking) edge[] node[pos=0.4,below,anchor=north east] {$\eta_2$} (idle);
        \end{tikzpicture}
        \caption{$\mathcal{N}_1\compose\mathcal{N}_2$ with $\mathcal{N}_1\stackrel{i}{\mapsto}\mathcal{N}_2$.}\label{fig:composed-inhibitory}
    \end{subfigure}
    \caption{Automaton with inhibitory synaptic connection: the external event $\sigma_2$ is replaced by the internal event $\eta_1$, as these events occur simultaneously.}\label{fig:automaton-inhibitory-synapse}
\end{figure}

If there is an inhibitory synaptic connection from neuron $\mathcal{N}_1$ to neuron $\mathcal{N}_2$, compactly denoted as $\mathcal{N}_1\stackrel{i}{\mapsto}\mathcal{N}_2$, then the transition $\sigma_2$ becomes internal, as neuron $\mathcal{N}_2$ is forced to transition from $i_2$ to $s_2$ whenever $\eta_1$ occurs. The resulting automata are given in Fig.~\ref{fig:automaton-inhibitory-synapse}.

\subsubsection{Bursting neurons}
When the neurons are burst excitable and rebound bursting, the interconnection ``rules'' presented above are still valid, although the spiking states are then replaced by bursting states. In case that the neuron is spike excitable but rebound bursting however, as is the case, e.g., in Fig.~\ref{fig:spiking_and_bursting}, some extra care is needed in forming the interconnections. The rebound bursting phenomenon typically only occurs if the neuron is inhibited \emph{sufficiently long}, which is usually the same timescale as the bursting itself. Hence, when $\mathcal{N}_1\stackrel{i}{\mapsto}\mathcal{N}_2$, a spike generated by neuron $\mathcal{N}_1$ will not trigger the bursting phenomenon in neuron $\mathcal{N}_2$. This means that essentially two timescales exist within a neuromorphic circuit: the \emph{slow} timescale, which captures mainly bursting behavior, and the \emph{fast} timescale, which captures mainly spiking behavior. Note that, when $\mathcal{N}_1\stackrel{e}{\mapsto}\mathcal{N}_2$ and $\mathcal{N}_1$ is bursting, $\mathcal{N}_2$ is spiking ``consecutively'' as long as $\mathcal{N}_1$ is bursting. Due to space limitations, we will not include the details of interconnecting rebound bursting neurons in this paper; this will be subject of future work, as it will require an additional ``discrete'' classification between the ``fast'' and ``slow'' timescales.

\subsection{DES of a network of neurons}\label{subsec:HCO}
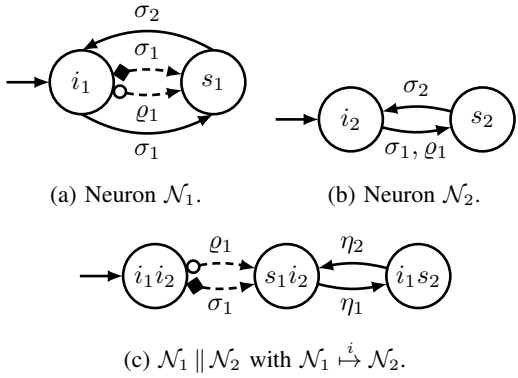
\begin{figure}[t!]
    \centering
    \begin{subfigure}{0.4\linewidth}
        \centering
        \begin{tikzpicture}
            \node[state] (idle) at (0,0) {$i_1$};
            \node[state] (spiking) at ($(idle)+(0:1.75cm)$) {$s_1$};
            \draw[<-] (idle) -- ++(180:1cm); 
            \path[{Turned Square}->,densely dashed] (idle) edge[bend left=5mm] node[midway, above] {$\sigma_1$} (spiking);
            \path[{Circle[open]}->,densely dashed] (idle) edge[bend right=5mm] node[midway, below] {$\varrho_1$} (spiking);
            \path[->] (spiking.90) edge[in=30, out=150] node[midway, above] {$\sigma_2$} (idle.90);
            \path[->] (idle.-90) edge[in=-150, out=-30] node[midway, below] {$\sigma_1$} (spiking.-90);
        \end{tikzpicture}
        \caption{Neuron $\mathcal{N}_1$.}
    \end{subfigure}~
    \begin{subfigure}{0.4\linewidth}
        \begin{tikzpicture}
            \node[state] (idle2) at (0,0) {$i_2$};
            \node[state] (spiking2) at ($(idle2)+(0:1.75cm)$) {$s_2$};
            \draw[<-] (idle2) -- ++(180:1cm);
            \path[->] (idle2) edge[bend right=5mm] node[midway, below] {$\sigma_1,\varrho_1$} (spiking2);
            \path[->] (spiking2) edge[bend right=5mm] node[midway, above] {$\sigma_2$} (idle2);
        \end{tikzpicture}
        \caption{Neuron $\mathcal{N}_2$.}
    \end{subfigure}
    \\[\belowdisplayshortskip]
    \begin{subfigure}[b]{0.8\linewidth}
        \centering
        \begin{tikzpicture}
            \node[state] (idle) at (0,0) {$i_1i_2$};
            \node[state] (spiking_idle) at ($(idle)+(0:1.75cm)$) {$s_1i_2$};
            \node[state] (idle_spiking) at ($(spiking_idle)+(0:1.75)$) {$i_1s_2$};
            \draw[<-] (idle) -- ++(180:1cm); 
            \path[{Turned Square}->,densely dashed] (idle) edge[bend right=5mm] node[midway, below] {$\sigma_1$} (spiking_idle);
            \path[{Circle[open]}->,densely dashed] (idle) edge[bend left=5mm] node[midway, above] {$\varrho_1$} (spiking_idle);
            \path[->] (spiking_idle) edge[bend right=5mm] node[midway, below] {$\eta_1$} (idle_spiking);
            \path[->] (idle_spiking) edge[bend right=5mm] node[midway, above] {$\eta_2$} (spiking_idle);
        \end{tikzpicture}
        \caption{$\mathcal{N}_1\compose\mathcal{N}_2$ with $\mathcal{N}_1\stackrel{i}{\mapsto}\mathcal{N}_2$.}\label{fig:composed-inhibitory-hco}
    \end{subfigure}
    \caption{Automata with mutual inhibitory synaptic connections.}\label{fig:automata_hco}
\end{figure}

We now construct the classical Half-Center Oscillator (HCO) and other ring oscillators as proposed in \cite{Huo_Forni_Sepulchre_2025}. We do this by adding mutual inhibitory connections to two neurons, i.e., we consider $\mathcal{N}_1\stackrel{i}{\mapsto}\mathcal{N}_2$ and $\mathcal{N}_2\stackrel{i}{\mapsto}\mathcal{N}_1$ for a pair of rebound spiking neurons. If we assume that we can only interact with neuron $\mathcal{N}_1$, the resulting automata are given in Fig.~\ref{fig:automata_hco}.
 
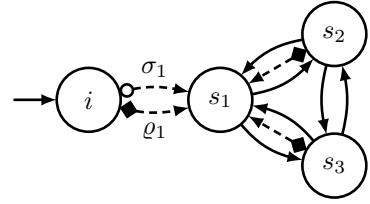
\begin{figure}[t!]
    \centering
    \begin{tikzpicture}
        \node[state] (i) at (0,0) {$i$};
        \node[state] (s1) at ($(i)+(0:1.75cm)$) {$s_1$};
        \node[state] (s2) at ($(s1)+(30:1.75cm)$) {$s_2$};
        \node[state] (s3) at ($(s1)+(-30:1.75cm)$) {$s_3$};
        \path[{Circle[open]}->, densely dashed] (i) edge[bend left=5mm] node[midway, above] {$\sigma_1$} (s1);
        \path[{Turned Square}->, densely dashed] (i) edge[bend right=5mm] node[midway, below] {$\varrho_1$} (s1);
        \path[->] (s1) edge[bend right=7mm] (s2);
        \path[->] (s1) edge[bend right=7mm] (s3);
        \path[->] (s2) edge[bend right=7mm] (s1);
        \path[->] (s3) edge[bend right=7mm] (s1);
        \path[->] (s2) edge[bend right=5mm] (s3);
        \path[->] (s3) edge[bend right=5mm] (s2);
        \path[{Turned Square}->,densely dashed] (s2) edge (s1);
        \path[{Turned Square}->,densely dashed] (s3) edge (s1);
        \draw[<-] (i) -- ++(180:1cm); 
    \end{tikzpicture}
    \caption{Automaton representation of an $N=3$ WTA network of rebound spiking neurons.}\label{fig:wta_3_neurons}
\end{figure}

Lastly, as a prelude to the next section, we will consider a network of $3$ post-inhibitory rebound spiking neurons with all-to-all inhibitory connections excluding self-inhibition. A couple of things change in this setting from the two-neuron case. Firstly, we consider the case where a small amount of noise is present in the membrane potential of each neuron, i.e., where the membrane potential of each neuron slightly fluctuates. This acts as a tie-breaker: suppose neuron $\mathcal{N}_1$ spikes, inhibiting neurons $\mathcal{N}_2$ and $\mathcal{N}_3$. After neuron $\mathcal{N}_1$ stops spiking, either $\mathcal{N}_2$ will spike first, inhibiting neuron $\mathcal{N}_3$, or neuron $\mathcal{N}_3$ spikes, inhibiting neuron $\mathcal{N}_2$. Which neuron successfully suppresses the other is essentially decided randomly. Suppose now that we are only allowed to interaction with neuron $\mathcal{N}_1$. The resulting automaton of this network is then given in Fig.~\ref{fig:wta_3_neurons}.

Note that we can, in this case, enforce the transition to $b_1$ from $b_2$ or $b_3$ by increasing the membrane potential ``externally,'' and recall that external transitions preempt internal ones. We will use these observations in the following to construct a general winner-take-all circuit for decision-making.

\section{Decision-making in Neuromorphic Circuits}\label{sec:wta}
A natural starting point for decision-making in neuromorphic circuits is the \emph{winner-take-all} (WTA) network \cite{Huo_Forni_Sepulchre_2025}, which is based on the winner-take-all principle \cite{Hopfield_1984,Maass_1999}. A WTA network is a network of $N\in\{2,3,\ldots\}$ neurons which have all-to-all inhibitory synaptic connections. The resulting behavior of a basic WTA circuit is summarized in Definition~\ref{def:WTA}.

\begin{definition}\label{def:WTA}
    Winner-take-all neuronal network properties:
    \begin{enumerate}[label=(\roman*)]
        \item\label{it:wta:win} Only one neuron can ``win'' (be on) at any given time.
        \item\label{it:wta:input} The neuron with the largest input wins. 
        \item\label{it:wta:consecutive} A neuron cannot win twice in a row.
    \end{enumerate}
\end{definition}

Some comments are in order. Firstly, we can view a neuron ``firing'' here as the representation of a decision that is made. Secondly, item~\ref{it:wta:input} considers inputs in the classical sense. As extensively discussed in Section~\ref{subsec:inputs}, the interpretation in the DES is slightly different. If we assume that some noise is present in the membrane potential of the neurons, this essentially results in a ``random'' \emph{other} neuron winning (so that~\ref{it:wta:consecutive} is not violated). Lastly, although item~\ref{it:wta:consecutive} is natural in a biological context (as we cannot have a stable tunable rhythm with a single neuron), the DES interpretation is that there are no self-loops. 

\subsection{Basic WTA automaton}
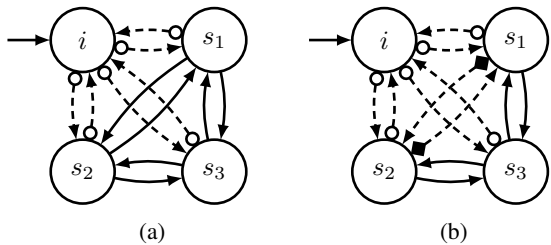
\begin{figure}[t!]
    \centering
    \begin{subfigure}{0.45\linewidth}
        % \centering
        \begin{tikzpicture}
            \node[state] (P0) at (0,0) {$i$};
            \node[state] (P1) at (1.75cm,0) {$s_1$};
            \node[state] (P2) at (0,-1.75cm) {$s_2$};
            \node[state] (P3) at (1.75cm,-1.75cm) {$s_3$};
            \draw[->] (P1) to[bend left=4mm] (P3);
            \draw[->] (P2) to[bend right=4mm] (P3);
            \draw[{Circle[open]}->,densely dashed] (P3) to[bend right=4mm] (P0);
            \draw[{Circle[open]}->,densely dashed] (P0) to[bend right=4mm] (P3);
            \draw[{Circle[open]}->,densely dashed] (P0) to[bend right=4mm] (P1);
            \draw[{Circle[open]}->,densely dashed] (P0) to[bend right=4mm] (P2);
            \draw[->] (P3) to[bend left=4mm] (P1);
            \draw[->] (P3) to[bend right=4mm] (P2);
            \draw[{Circle[open]}->,densely dashed] (P2) to[bend right=4mm] (P0);
            \draw[{Circle[open]}->,densely dashed] (P1) to[bend right=4mm] (P0);
            \draw[->] (P2) to[bend right=4mm] (P1);
            \draw[->] (P1) to[bend right=4mm] (P2);
            \draw[<-] (P0) -- ++(180:1cm);
        \end{tikzpicture}
        \caption{}\label{fig:automaton_undirected}
    \end{subfigure}
    \begin{subfigure}{0.45\linewidth}
        % \centering
        \begin{tikzpicture}
            \node[state] (P0) at (0,0) {$i$};
            \node[state] (P1) at (1.75cm,0) {$s_1$};
            \node[state] (P2) at (0,-1.75cm) {$s_2$};
            \node[state] (P3) at (1.75cm,-1.75cm) {$s_3$};
            \draw[->] (P1) to[bend left=4mm] (P3);
            \draw[->] (P2) to[bend right=4mm] (P3);
            \draw[{Circle[open]}->,densely dashed] (P3) to[bend right=4mm] (P0);
            \draw[{Circle[open]}->,densely dashed] (P0) to[bend right=4mm] (P3);
            \draw[{Circle[open]}->,densely dashed] (P0) to[bend right=4mm] (P1);
            \draw[{Circle[open]}->,densely dashed] (P0) to[bend right=4mm] (P2);
            \draw[->] (P3) to[bend left=4mm] (P1);
            \draw[->] (P3) to[bend right=4mm] (P2);
            \draw[{Circle[open]}->,densely dashed] (P2) to[bend right=4mm] (P0);
            \draw[{Circle[open]}->,densely dashed] (P1) to[bend right=4mm] (P0);
            \draw[{Turned Square}->,densely dashed] (P2) to[bend right=4mm] (P1);
            \draw[{Turned Square}->,densely dashed] (P1) to[bend right=4mm] (P2);
            \draw[<-] (P0) -- ++(180:1cm);
        \end{tikzpicture}
        \caption{}\label{fig:automaton_directed}
    \end{subfigure}
    \caption{Automaton representation of the discrete dynamics of an $N=3$ WTA network. The state $i$ represents the neurons being idle, while $s_1$, $s_2$, and $s_3$ represent each respective neuron firing. In~\ref{fig:automaton_directed}, we add $\mathcal{N}_1\stackrel{e}{\mapsto}\mathcal{N}_3$ and $\mathcal{N}_2\stackrel{e}{\mapsto}\mathcal{N}_3$ to enforce a partial temporal ordering.}
\end{figure}

Using the insight obtained in Section~\ref{subsec:HCO}, it is straightforward to derive the following proposition.
\begin{proposition}\label{prop:wta}
    The discrete dynamics of a winner-take-all network of $N$ post-inhibitory rebound spiking neurons with no excitatory synapses can be captured in an $N+1$ automaton with all-to-all transitions excluding self-loops.
\end{proposition}
Note that, indeed, this network satisfies items~\ref{it:wta:win}-\ref{it:wta:consecutive} in Definition~\ref{def:WTA}.
Thus, as was already mentioned in the above, we can model a WTA-network as an automaton by simply assigning a state in the model that represents a single neuron ``winning,'' i.e., firing. An example of a DES models that represent the behavior of Definition~\ref{def:WTA} with $N=3$ neurons is given in Fig.~\ref{fig:automaton_undirected}. Note that the resulting model is fairly tractable: the number of states is linear in the number of neurons. We will exploit this fact in Section~\ref{subsec:realization}.

\subsection{Adding excitatory synaptic connections}
As mentioned earlier, adding (slow) excitatory synapses enforces a particular ordering in the firing order of the neurons. Consequently, it ``disables'' transitions from the DES model. In this way, we can restrict the set of possible transitions neuron $i\in\{2,3,\ldots,N\}$ ``internally'' (i.e., without external inputs). Essentially, this causes some transitions to become external; we can still transition from $b_2$ to $b_1$, but it requires an external current being applied to neuron $\mathcal{N}_1$. This also implies that, when every neuron has exactly one excitatory synapse connecting it to another neuron, the firing order is a fixed (temporal) pattern.

\subsection{Towards a realization theory}\label{subsec:realization}
One can easily verify that, with these elementary building blocks, any automaton without self-loops can be built and/or implemented as a neuromorphic circuit. Such a circuit can be constructed as follows:
\begin{enumerate*}[label=(\roman*)]
    \item For each state, add a neuron in a WTA network.
    \item For each transition, add a slow excitatory synapse to ``force'' this transition internally.
\end{enumerate*}
Clearly, when a small amount of noise is used as a tie-breaker when two (or no) excitatory synapses originate from a single neuron, such a circuit will mimic the behavior of the DES: the noise will ``select'' the next state randomly (from the set of connected neurons in case multiple excitatory synaptic connections). These principles where already suggested in the work Huo et al. \cite{Huo_Forni_Sepulchre_2025}, which motivated the question of the paper.

\section{Conclusions and discussion}\label{sec:conclusions}
In this paper, we have motivated the value of discrete-event models of neuromorphic circuits and shown how to construct DES models that capture the discrete behavior in terms of the sequence of events. This approach is a complementary approach to the biophysical modeling as (conductance-based) differential equations. Indeed, from the conductance-based model, it is significant effort to grasp the ordering of the events, whereas for a DES model, it is impossible to do regulation properly if timings are not present. However, as timings can be regulated using neuromodulation, we believe that both models serve complementary objectives.

In our DES models, pairs of mixed conductances add states whereas synaptic interconnections add internal transitions. We have also seen that the WTA architecture results in \emph{tractable} models. Lastly, we shortly demonstrated how the models and observations in this paper can be used to construct a neuromorphic circuit which represents or mimics the behavior of a specific automaton.

We envision that, e.g., in the field of robotics, adding decision-making circuitry enables the design of more complicated interactive behaviors. In particular, when parts of the low-level control loops are implemented using half-center oscillators, such decision-making capabilities can be added to the circuitry by using the same elementary building blocks. The framework of (timed or untimed) DES can be used to analyze such circuits and to design behaviors, while the continuous-time conductance-based models can be used to tackle the regulation problem.
\bibliography{library}

\end{document}